\documentclass[conference]{IEEEtran} 

\let\labelindent\relax
\usepackage{enumitem}

\usepackage{flushend}

\usepackage{amsmath}          
\usepackage{amsfonts}         
\usepackage{graphicx} 
\usepackage{subfigure}        
\usepackage{epsfig}
\usepackage{epstopdf}

\usepackage[font=small,aboveskip=10pt]{caption}
\usepackage{color}			  
\usepackage[nolist]{acronym}					
\usepackage{multirow}
\usepackage{algorithmic,algorithm}
\usepackage[noadjust]{cite}       

\usepackage{xspace}
\definecolor{correction_colour}{rgb}{0,0,0} 
\definecolor{correction}{rgb}{0,0,0} 

\usepackage{todonotes}

\begin{document}

\bstctlcite{IEEEexample:BSTcontrol}

\title{Implementing NB-IoT in Software - \\Experiences Using the srsLTE Library}

\author{\IEEEauthorblockN{Andr\'{e} Puschmann,
Paul Sutton,
Ismael Gomez}

\IEEEauthorblockA{Software Radio Systems Ltd., Cork, Ireland\\
Email: \{andre, paul, ismael\}@softwareradiosystems.com}}

\maketitle

\begin{abstract}
NB-IoT is the 3GPP standard for machine-to-machine communications, recently finalized
within LTE release 13. This article gives a brief overview about this new LTE-based
radio access technology and presents a implementation developed using the srsLTE software radio suite.
We also carry out a performance study in which we compare a theoretical analysis
with experimental results obtained in our testbed.
Furthermore, we provide some interesting details and share our experience in exploring
one of the worldwide first commercial NB-IoT deployments.

\end{abstract}
\begin{acronym}
%
%
%
%
%
\acro{3gpp}[3GPP]{3\textsuperscript{rd} Generation Partnership Program}
\acro{acb}[ACB]{Adaptive Cognitive Beamforming}
\acro{aoa}[AoA]{Angle of Arrival}
\acro{aod}[AoD]{Angle of Departure}
\acro{amdf}[AMDF]{Average Magnitude Difference Function}
\acro{arp}[ARP]{Antenna Radiation Pattern}
\acro{asp}[ASP]{Antenna Scan Period}
\acro{awgn}[AWGN]{Additive White Gaussian Noise}
\acro{bler}[BLER]{BLock Error Rate}
\acro{cact}[CACT]{Channel Availability Check Time}
\acro{ccc}[CCC]{Common Control Channel}
\acro{cdi}[CDI]{Channel Direction Indicator}
\acro{cdif}[CDIF]{Cumulative Difference Histogram}
\acro{csi}[CSI]{Channel State Information}
\acro{cs-rs}[CS-RS]{Cell Specific Reference Signal}
\acro{cqi}[CQI]{Channel Quality indicator}
\acro{cr}[CR]{Cognitive Radio}
\acro{crs}[CRS]{Cell-Specific Reference Signal}
\acro{cran}[C-RAN]{Cloud Radio Access Network}
\acro{crs}[CRS]{Common Reference Signal}
\acro{cw}[CW]{Constant Wave}
\acro{dfs}[DFS]{Dynamic Frequency Selection}
\acro{dl}[DL]{Downlink}
\acro{dm-rs}[DM-RS]{Demodulation Reference Signal}
\acro{dod}[DoD]{Department of Defense}
\acro{dsa}[DSA]{Dynamic Spectrum Access}
\acro{dthres}[DT]{Detection Threshold}
\acro{eic}[EIC]{Effective Interference Channel}
\acro{eip}[EIP]{Expected Interference Period}
\acro{eirp}[EIRP]{Equivalent Isotropic Radiated Power}
\acro{elint}[ELINT]{Electronic Intelligence}
\acro{em}[EM]{Expectation Maximization}
\acro{enb}[eNB]{evolved Node Basestation}
\acro{esc}[ESC]{Environmental Sensing Capability}
\acro{ew}[EW]{Electronic Warfare}
\acro{fcc}[FCC]{Federal Communications Commission}
\acro{fdd}[FDD]{Frequency Division Duplexing}
\acro{fdr}[FDR]{Frequency Dependent Rejection}
\acro{fss}[FSS]{Fixed Satellite Services}
\acro{gaa}[GAA]{General Authorized Access}
\acro{gldb}[GL-DB]{Geolocation Database}
\acro{hmm}[HMM]{Hidden Markov Model}
\acro{inr}[INR]{Interference to Noise Ratio}
\acro{ipm}[IPM]{Intra-Pulse Modulation}
\acro{itm}[ITM]{Irregular Terrain Model}
\acro{itu-r}[ITU-R]{International Telecommunication Union - Radiocommunication sector}
\acro{iw}[IW]{Interweave}
\acro{lfm}[LFM]{Linear Frequency Modulation}
\acro{los}[LOS]{Line Of Sight}
\acro{lsa}[LSA]{Licensed Shared Access}
\acro{lte}[LTE]{Long Term Evolution}
\acro{lte-a}[LTE-A]{\ac{lte}-Advanced}
\acro{mac}[MAC]{Media Access Control}
\acro{mcs}[MCS]{Modulation and Coding Scheme}
\acro{mimo}[MIMO]{Multiple Input Multiple Output}
\acro{mse}[MSE]{Minimum Squared Error}
\acro{mu-mimo}[MU-MIMO]{Multi-User MIMO}
\acro{mui}[MUI]{Multi-User Interference}
\acro{nlos}[NLOS]{Non-Line Of Sight}
\acro{ntia}[NTIA]{National Telecommunications and Information Administration}
\acro{ofcom}[OFCOM]{Office of Communications}
\acro{ofdm}[OFDM]{Orthogonal Frequency Division Multiplexing}
\acro{osa}[OSA]{Opportunistic Spectrum Access}
\acro{pa}[PA]{Pulse Amplitude}
\acro{pal}[PAL]{Priority Access License}
\acro{pamp}[PAmp]{Pulse Amplitude}
\acro{parp}[PARP]{Perceived Antenna Radiation Pattern}
\acro{pc}[PC]{Pulsed Carrier}
\acro{pdw}[PDW]{Pulse Description Word}
\acro{pf}[PF]{Pulse Frequency}
\acro{pfreq}[PFreq]{Pulse Frequency}
\acro{phc}[PhC]{Phantom Cell}
\acro{phy}[PHY]{Physical} 
\acro{pm}[PM]{Phase Modulation}
\acro{pmi}[PMI]{Precoding Matrix Indicator}
\acro{prf}[PRF]{Pulse Repetition Frequency}
\acro{pri}[PRI]{Pulse Repetition Interval}
\acro{psnr}[PSNR]{Peak Signal-to-Noise Ratio}
\acro{pt}[PT]{Pulse Train}
\acro{pu}[PU]{Primary User}
\acro{pw}[PW]{Pulse Width}
\acro{qp}[QP]{Quiet Period}
\acro{rat}[RAT]{Radio Access Technology}
\acro{rb}[RB]{Resource Block}
\acro{rf}[RF]{Radio Frequency}
\acro{rmse}[RMSE]{Root Mean Squared Error}
\acro{roc}[ROC]{Receiver Operating Characteristic}
\acro{roi}[ROI]{Region of Interest}
\acro{rs}[RS]{Reference Signalling}
\acro{rsrp}[RSRP]{Reference Signal Received Power}
\acro{rx}[Rx]{Receiver}
\acro{sa}[SA]{Spectrum Access}
\acro{sas}[SAS]{Spectrum Access System}
\acro{sdr}[SDR]{Software Defined Radio}
\acro{snr}[SNR]{Signal to Noise Ratio}
\acro{sinr}[SINR]{Signal to Interference plus Noise Ratio}
\acro{sl}[SL]{System Level}
\acro{ss}[SS]{Spectrum Sensing}
\acro{su}[SU]{Secondary User}
\acro{tdd}[TDD]{Time Division Duplexing}
\acro{tbs}[TBS]{Transport Block Size}
\acro{twdr}[TWDR]{Terminal Doppler Weather Radar}
\acro{toa}[TOA]{Time Of Arrival}
\acro{tot}[ToT]{Time on Target}
\acro{ts}[TS]{Temporal Sharing}
\acro{tti}[TTI]{Transmission Time Interval}
\acro{tvws}[TVWS]{TV White Space}
\acro{tx}[Tx]{Transmit}
\acro{ue}[UE]{User Equipment}
\acro{ul}[UL]{Uplink}
\acro{wlan}[WLAN]{Wireless Local Area Network}
\acro{ws}[WS]{White Space}
\acro{zf}[ZF]{Zero-Forcing}
\acro{zfbf}[ZFBF]{Zero-Forcing Beam Forming}
\end{acronym}
\newcommand{\Nstates}{N_\textup{states}}
\newcommand{\Nparams}{N_\textup{params}}

\acresetall 

\section{Introduction}
\label{sec:introduction}

Narrowband Internet of Things (NB-IoT) has been standardized by 3GPP in Release 13
as a new air interface with the aim to make LTE suitable for low-cost
massive machine-type communication (m-MTC). NB-IoT is heavily based on LTE but
makes a number simplifications and optimizations in order to reduce device costs,
minimize power consumption and to make it work in unfavourable radio conditions.

This work presents an implementation of the NB-IoT standard, developed using
the srsLTE software radio suite within the FP7 FLEX-IoT project.
Since 2008, the Future Internet Research and Experimentation~(FIRE) initiative
has bridged the gap between visionary research and large-scale experimentation on
new networking and service architectures and paradigms. The FLEX~(FIRE LTE testbeds
for open EXperimentation) project provides an open and operational LTE experimental
facility based on a combination of configurable commercial equipment,
configurable core network software, open-source components, and sophisticated 
emulation and mobility functionalities. Under the FLEX-IoT project,
we have integrated our high performance software radio UE~(srsUE) as an extension
to the FLEX testbed facilities and extended the FLEX testbed facilities to include
the NB-IoT features of 3GPP LTE release 13 for LPWAN.
For an excellent overview about NB-IoT in general and it's differences to LTE in
particular the interested reader is referred to \cite{Wang2016} and \cite{rohde-paper-nbiot}.

In this report we provide a brief overview about NB-IoT and describe
the available channels and signals in Section \ref{sec:nb-iot}. We then discuss
the architecture of srsLTE and how the new NB-Iot components have been integrated into it.
The main contributions of this article is a performance analysis carried out in
Section \ref{sec:perf-eval}. We study the achievable throughput in the downlink
and compare results obtained from practical experiments with those from a theoretical analysis.
In Section \ref{sec:mwc} we share our experience in exploring one of the first
commercial NB-IoT deployments.
Finally, we conclude the paper in Section \ref{sec:conclusion}.

\section{NB-IoT Overview}
\label{sec:nb-iot}

This section provides a brief overview about NB-IoT, it's transmission schemes, downlink~(DL)
and uplink~(UL) channels and signals and the different deployment schemes.

\subsection{Transmission schemes, time structure and deployment options}
NB-IoT uses the same downlink transmission scheme like normal LTE with OFDMA
with a subcarrier spacing of 15~kHz. 12 subcarriers are combined to form a 
single carrier with a bandwidth of 180 kHz, i.e., NB-IoT only uses one 
physical resource block~(PRB) instead of up to 100~PRBs.
In the time domain, it also uses the same frame structure with frames of 10~ms length.
Each frame consists of 10 subframes of 1~ms length and each subframe consists of
2 slots with a length of 7~OFDM symbols each. NB-IoT only defines the normal cyclic prefix.
In addition to that, NB-IoT also defines the concept of so-called hyper frames.
Similar to the system frame number~(SFN), the hyper frame number~(HFN) is also a
ten-bit wide number that is incremented whenever the SFN wraps around, i.e., every 10240~ms. 

In the uplink, NB-IoT provides two options: single and multi-tone transmissions.
In the latter case, which is the default option, it uses the same SC-FDMA scheme
with a 15~kHz subcarrier spacing and a total bandwidth of 180~kHz as LTE.

NB-IoT offers multiple deployment options. It can be deployed in standalone mode, i.e.,
without any pre-existing LTE network, or together with an already deployed LTE network.
The latter has the advantage that existing sites can be used to roll out an IoT network
without heavy investments in new hardware. In this case, the NB-IoT
carrier can be placed inside one of the subcarriers of an existing LTE cell or in the
guardbands of a cell. In Release 13, 3GPP has also standardized a new UE device category
for NB-IoT, called “Cat-NB1”.

\subsection{Downlink}

\begin{figure}
	\begin{center}
		\includegraphics[width=1.0\columnwidth]{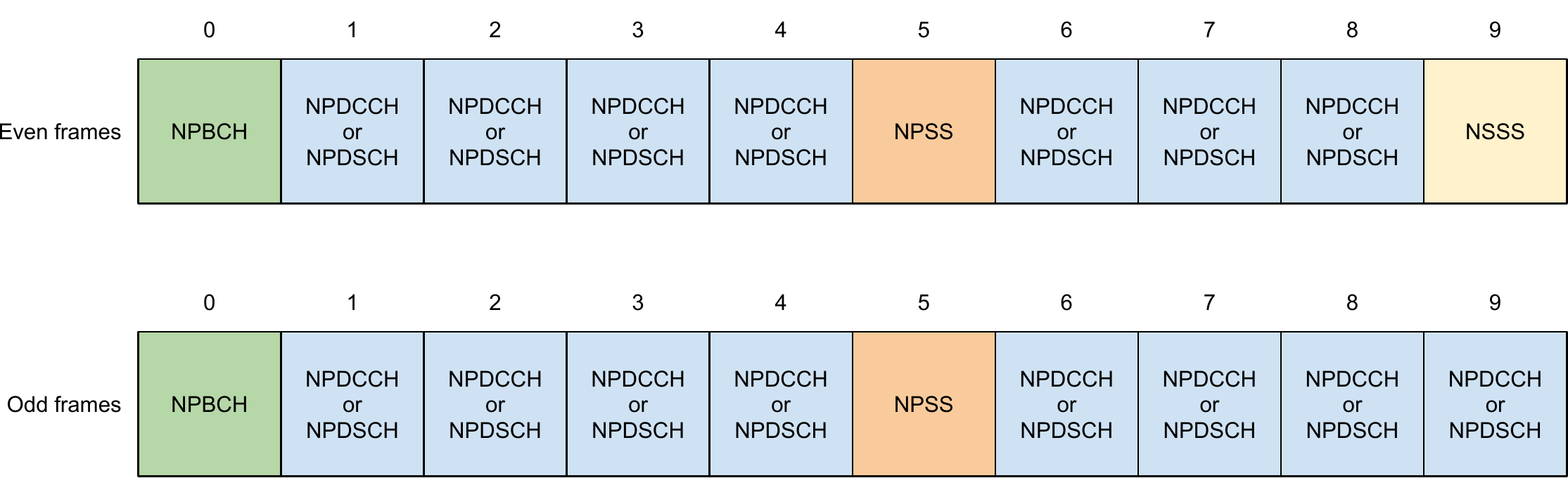}
		\caption{NB-IoT downlink frame structure and time multiplexing (reproduced from \cite{Wang2016}).}
		\label{nbiot-chan-multiplexing}
	\end{center}	
\end{figure}

Release 13 of the 3GPP standard defines six physical DL channels/signals for NB-IoT, all of which
will be briefly described below~\cite{ts36211}. Because NB-IoT only uses a single PRB, the channels are only
multiplexed in time-domain. Figure~\ref{nbiot-chan-multiplexing} depicts how all DL signals are 
transmitted over the length of two frames.

\subsubsection{Narrowband Primary Synchronization Signal~(NPSS)}

The NPSS, as well as the NSSS described below, are primarily used for initial
cell search and for acquiring time and frequency synchronization. The NPSS is transmitted
in every frame in subframe number 5. It only uses 11 of the 14 available OFDM symbols of the
subframe to protect a potential LTE transmission in the same subframe. The signal itself
consists of a single length-11 Zadoff-Chu~(ZC) sequence that is either directly mapped to
the 11 lowest subcarriers (the 12th subcarrier is empty in the NPSS) or is inverted before
the mapping process. After successful detection of the NPSS, a NB-IoT UE is able to determine
the frame boundaries of a DL transmission.

\subsubsection{Narrowband Secondary Synchronization Signal~(NSSS)}
The NSSS is transmitted every 20~ms in every even-numbered frame in subframe number 9.
The signal  consists of a length-132 sequence that is generated from a scrambling sequence
and a length-132 ZC sequence. The NSSS only carries the cell ID.

\subsubsection{Narrowband Reference Signal~(NRS)}
The NRS is required by the UE to successfully demodulate any of the data or control channels.
For this purpose, the eNB broadcasts a known pilot signal that can be used to estimate the DL channel.
With this knows reference, the UE tries to correct the effects of the channel before demodulating
the signal. The NRS is transmitted in every subframe carrying either the NBPCH, NPDCCH or the NPDSCH.
It is distributed over 4~OFDM symbols (the last and second last symbol of each slot) and uses a total
of 8 resource elements~(RE) per antenna port.

\subsubsection{Narrowband Physical Broadcast Channel~(NPBCH)}
The NPBCH is the first physical channel that a UE attempts to receive and demodulate after
synchronizing with a cell. The NPBCH conveys important information that is encoded in the Narrowband master
information block~(MIB-NB), such as the SFN (the four most significant bits), the HFN (the two least significant bits),
system information block~(SIB) scheduling information and the NB-IoT operation mode. 
In order to allow successful decoding of the MIB-NB also in extraneous radio conditions, it is transmitted in the
first subframe of every ever frame over a total period of 640ms, i.e., over 64 frames.
For this purpose, the MIB-NB signal is split into 8 blocks, each of which is repeated 8 times.
The NPBCH is QPSK modulated and uses a tail-biting convolution encoding~(TBCC).

\subsubsection{Narrowband Physical Downlink Control Channel~(NPDCCH)}
The NPDCCH is the only control channel defined for NB-IoT, i.e., there is now uplink counterpart
like in normal LTE. It carries the DL and UL scheduling grants,
acknowledgment information for UL transmissions, paging indication, system information update
and random access response (RAR) scheduling information. For NB-IoT only three DL control information~(DCI) formats
are specified and only three possible positions exist in the NPDCCH. This simplifies the decoding overhead for the
UE significantly compared to LTE. Like the NPBCH, the NPDCCH is also QPSK modulated and protected using TBCC.

\subsubsection{Narrowband Physical Downlink Shared Channel~(NPDSCH)}
The NPDSCH is the main data channel and carries SIBs, upper layer data, and random access response~(RAR) messages.

\subsection{Uplink}

The 3GPP set of specifications defines two uplink channels, the Narrowband Physical Random Access Channel~(NPRACH)
and the Narrowband Physical Uplink Shared Channel~(NPUSCH). It is interesting to note that there is no equivalent
to the Physical Uplink Control Channel~(PUCCH) of LTE in NB-IoT. In addition to that, NB-IoT defines the Narrowband
demodulation reference signal~(DMRS).

\subsubsection{Narrowband Physical Random Access Channel (NPRACH)}
As in normal LTE, the NPRACH is used by an UE to signal the eNB that it wants to establish a connection with it.
The attachment procedure is a contention-based access method in which the UE transmits a random access preamble.
One such preamble consists of four symbol groups each of which is transmitted on a different subcarrier.
Each symbol group in turn consists of the cyclic prefix~(CP) plus 5~symbols, with a total length of either 1.4~ms or 1.6~ms,
depending on the format defined by the eNB. In order to increase the coverage the eNB may request a UE to repeat
the preamble transmission up to 128 times. The subcarriers used to transmit the NPRACH follow a certain hopping
pattern that depends on a randomly selected initialization value.

\subsubsection{Narrowband Physical Uplink Shared Channel (NPUSCH)}
The only means to communicate data from the UE to the eNB, except for the random access, is over the NPUSCH.
There exist two formats, NPUSCH format 1 is used to carry user data, whereas format 2 is used to carry
uplink control information.
Upon reception of the preamble, the eNB responds with the random access response~(RAR) which initiates
the transmission of the first scheduled transmission by the UE~(Msg3). After the contention resolution
sent from the eNB, the NB-IoT random access procedure is complete.

\subsubsection{Narrowband demodulation reference signal~(DMRS)}
In order for the eNB to be able to estimate the UL channel, the UE transmits a demodulation reference signal~(DMRS)
that is multiplexed with the actual NPUSCH symbols. Depending on the NPUSCH format, i.e. either format 1 or format 2,
one or three symbols per SC-FDMA slot are used to transmit the DMRS. For example, for the case of format 1 NPUSCH
with a subcarrier spacing of 15 kHz, the 4th symbol of every slots contains the DMRS symbols in all allocated subcarriers.

\subsubsection{Uplink Baseband Signal Generation}
In order to generate the SC-FDMA baseband signal for the NB-IoT UL it is important to note an interesting
difference to normal LTE. In fact, each modulation symbol (both data and reference symbols) for
all UL transmissions undergo a phase rotation that depends on the position of the particular symbol. This reduces the peak-to-average power ratio~(PAPR) because it effectively yields a $\pi/2$-BPSK or $\pi/4$-QPSK modulation scheme with phase continuity between symbols, i.e., no zero-crossing in the constellation diagram between symbols.

\section{srsLTE}
\label{sec:srslte}

srsLTE is a high-performance open-source software radio LTE physical layer library~\cite{srslte}.
It is specifically designed to support new air interface modifications and enhancements using
a modular architecture with clean inter-component interfaces.

\subsection{Library Structure}

\begin{figure}
	\begin{center}
		\includegraphics[width=1.0\columnwidth]{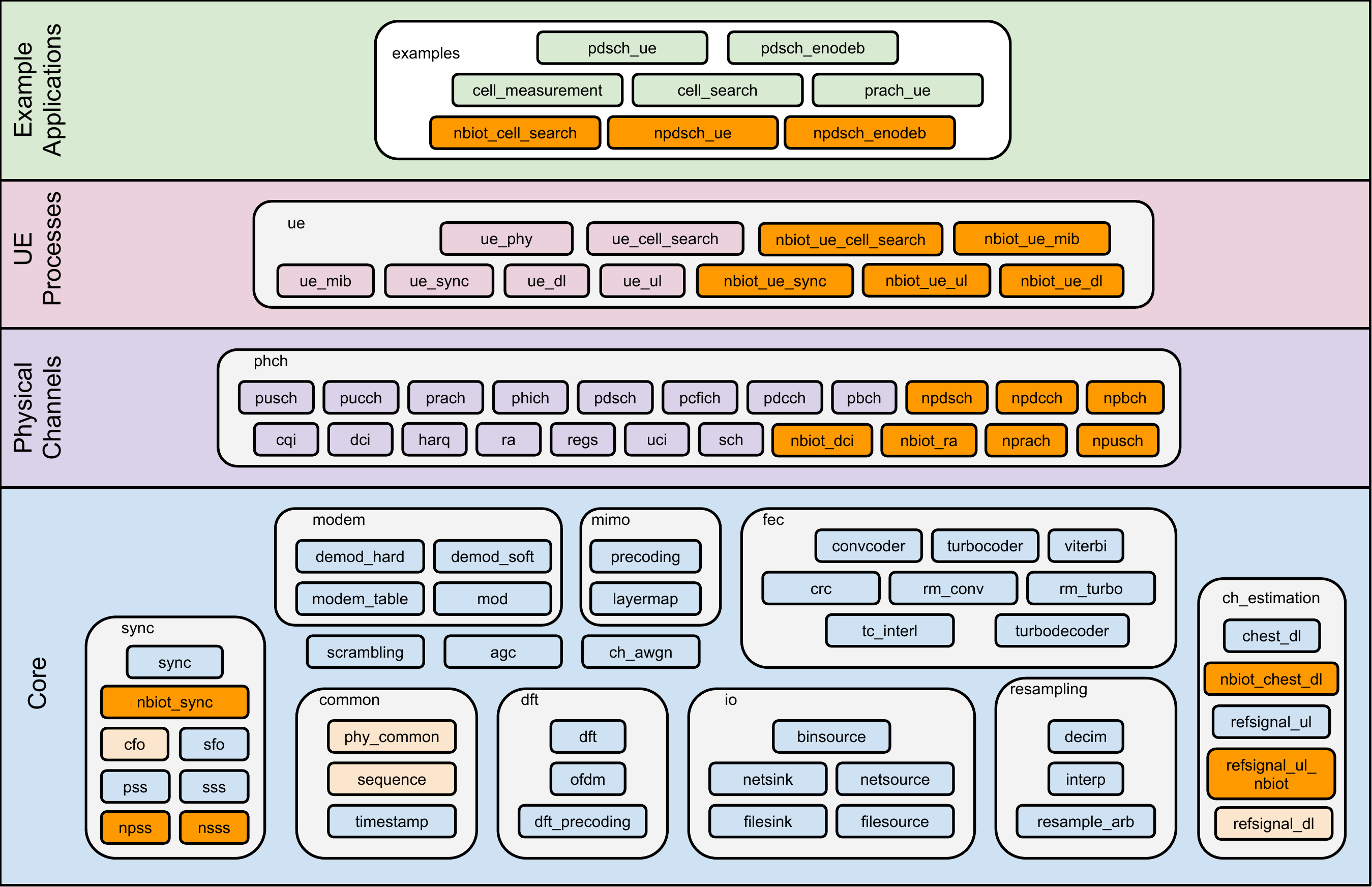}
		\caption{srsLTE's modular architecture with new NB-IoT components.}
		\label{fig:srslte}
	\end{center}
\end{figure}

Figure \ref{fig:srslte} shows the modular architecture of the srsLTE PHY-layer library.
Functional modules are organized into a hierarchy from elementary DSP components at the bottom
to physical channels, processes and example applications at the top. 

\subsection{NB-IoT Extension}

As part of the FLEX-IoT project, srsLTE has been extended to support the new NB-IoT radio access technology.
In doing so, we leveraged from the great modularity and extensibility of the library and consequently followed
the same approach for the NB-IoT physical layer components, including all DL/UL channels and
signals defined in the standard. All orange blocks in Figure \ref{fig:srslte} have been added
to srsLTE during the course of the NB-IoT extension.

srsLTE provides two example applications that act as PHY-only NB-IoT eNodeB and UE, similar to the already
existing counterparts for LTE. The \texttt{npdsch\_enodeb} example acts as an eNodeB by accepting data traffic
on a TCP socket and transmitting it using the NB-IoT downlink physical channels. The number of subframes
and the Transport Block Size~(TBS) used for the downlink transmissions can be dynamically changed by the
user through the command line.

The \texttt{npdsch\_ue} example acts as the counterpart and provides a PHY-only UE that
receives and decodes the DL transmission of the eNB.

\section{Performance Evaluation}
\label{sec:perf-eval}

This section evaluates the achievable downlink throughput in NB-IoT.
We carry out a theoretical analysis first before comparing the results with
experimental data obtained from our NB-IoT implementation.

\subsection{Theoretical Analysis}

\begin{figure*}
	\centering
	\subfigure[Unacknowledged mode (UM).\label{fig:tp_um}]{\includegraphics[width=1.0\columnwidth]{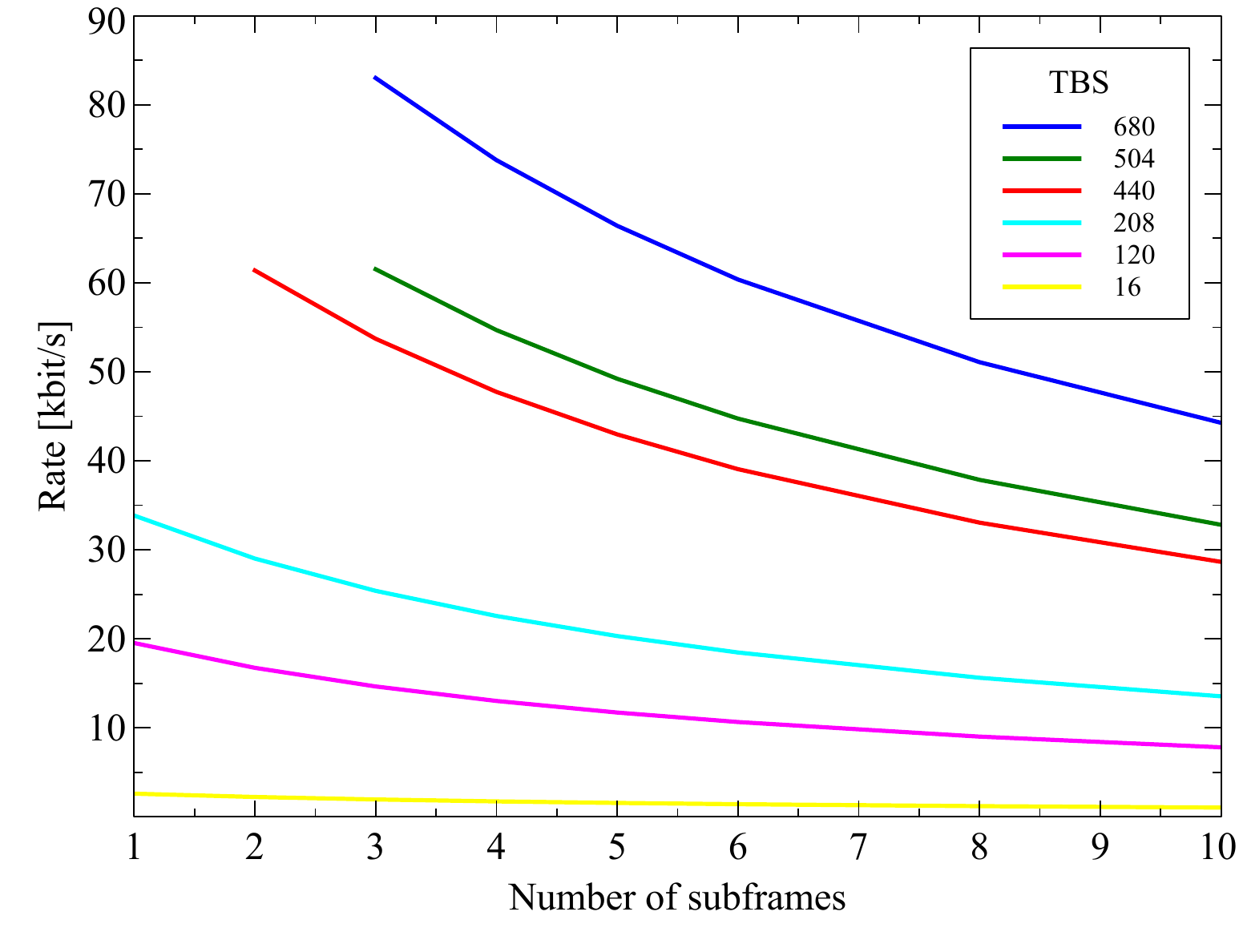}}\quad
	\subfigure[Acknowledged mode (AM).\label{fig:tp_am}]{\includegraphics[width=1.0\columnwidth]{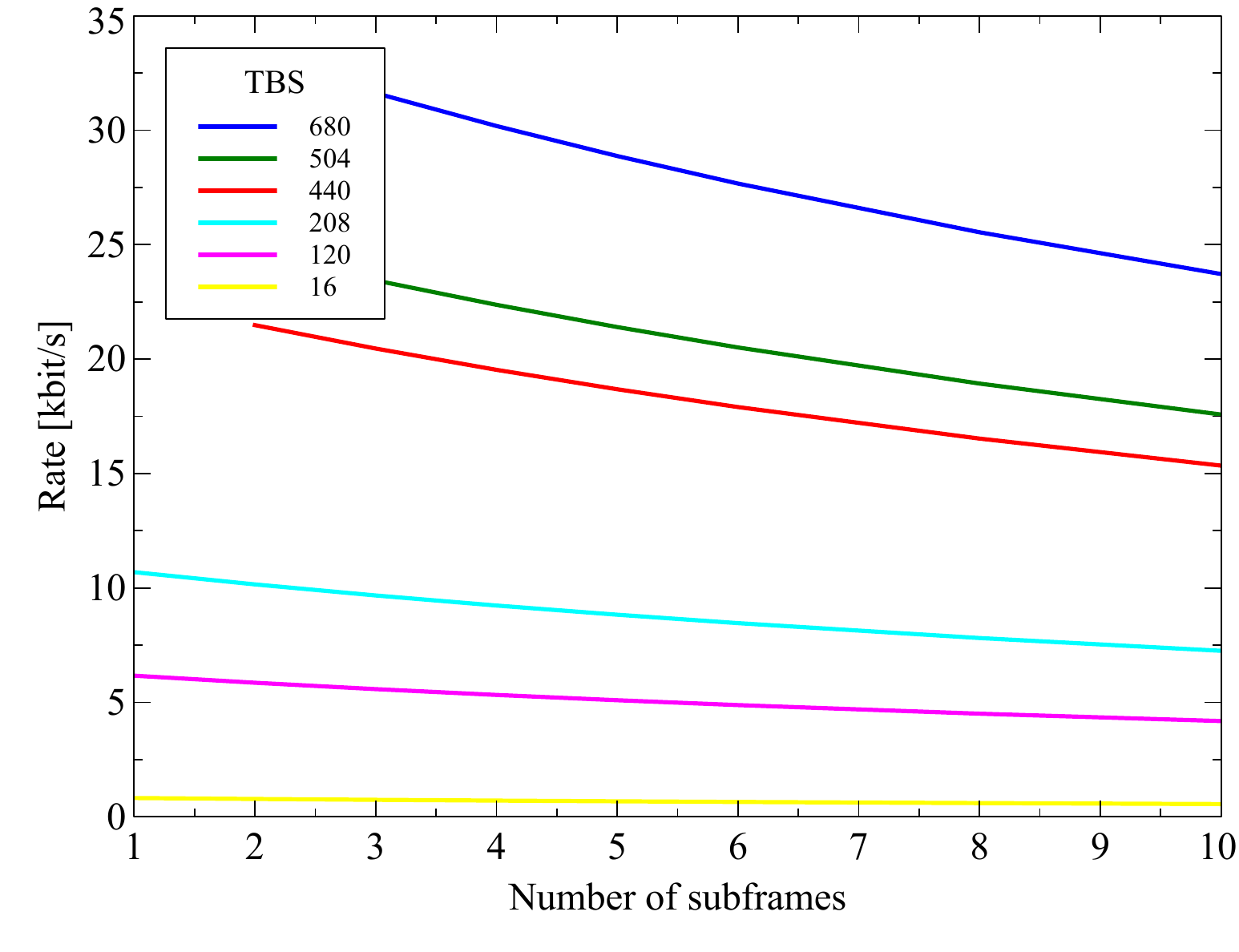}}
	\caption{Maximum achievable DL rate for a single UE}
\end{figure*}

As for normal LTE, data to be transmitted at the physical layer in NB-IoT
needs to have a size of a so-called transport block~(TB). The transport block size~(TBS)
for the DL ranges between 16 and 680~bits. Depending the configuration, a single
TB may be carried by one or more subframes. Transporting the largest TBS of 680~bits
requires at least 3 subframes. Therefore, the often stated peak data rate in the DL is 226~kbit/s.
This value, however, is only of a theoretical nature as it doesn't include the overhead
associated with any NPDSCH transmission (except for SIB transmissions). This overhead
includes at least one subframe for the NPDCCH to carry the DL grant as well as a minimum of
4~empty subframes that allow the UE to receive and decode the grant and to prepare for the
reception of the actual NPDSCH. In total, at least 8~ms are needed to transmit the full TB
carrying 680~bits. Please note that in NB-IoT there can only be a single hybrid
automatic repeat request~(HARQ) process running at the same time. Therefore, interleaved
NPDCCH/NPDSCH transmissions are not possible. We also assume only a single transmission
of each NPDSCH, i.e., now repetitions. This results in a maximum achievable DL data
rate in unacknowledged mode~(UM) of 85~kbit/s. Depending on the actual scheduling constraints
of the eNB and the number of subframes and repetitions for the NPDSCH, this number may be
much lower in reality. Figure \ref{fig:tp_um} shows the maximum achievable DL rates for
a single UE in UM for a number of TBS sizes as a function of the number of NPDSCH subframes.
Note that because not all TBS sizes are possible for every number of subframes, we selected the
closest possible value for each configuration in order to make the results comparable to each other.

In acknowledged mode (AM), the actual achievable rate reduces even further than in UM because there is an extra gap between the NPDSCH (data) and the NPUSCH (acknowledgement) transmission of at least 12 subframes. Therefore, even in the best case, at least 21ms are required to transfer 680~bits worth of data, resulting in a maximum rate of 32,38 kbit/s. This again assumes a single NPDCCH subframe for the grant, a 4ms gap, 3 subframes for the NPDSCH, a 12ms gap and one subframe carrying the NPUSCH with the acknowledgement. Figure \ref{fig:tp_am} plots the maximum achievable DL rate for a single UE in AM for a number of TBS as a function of the number of NPDSCH subframes.

\subsection{Experimental Results}

\begin{figure}
	\begin{center}
		\includegraphics[width=1.0\columnwidth]{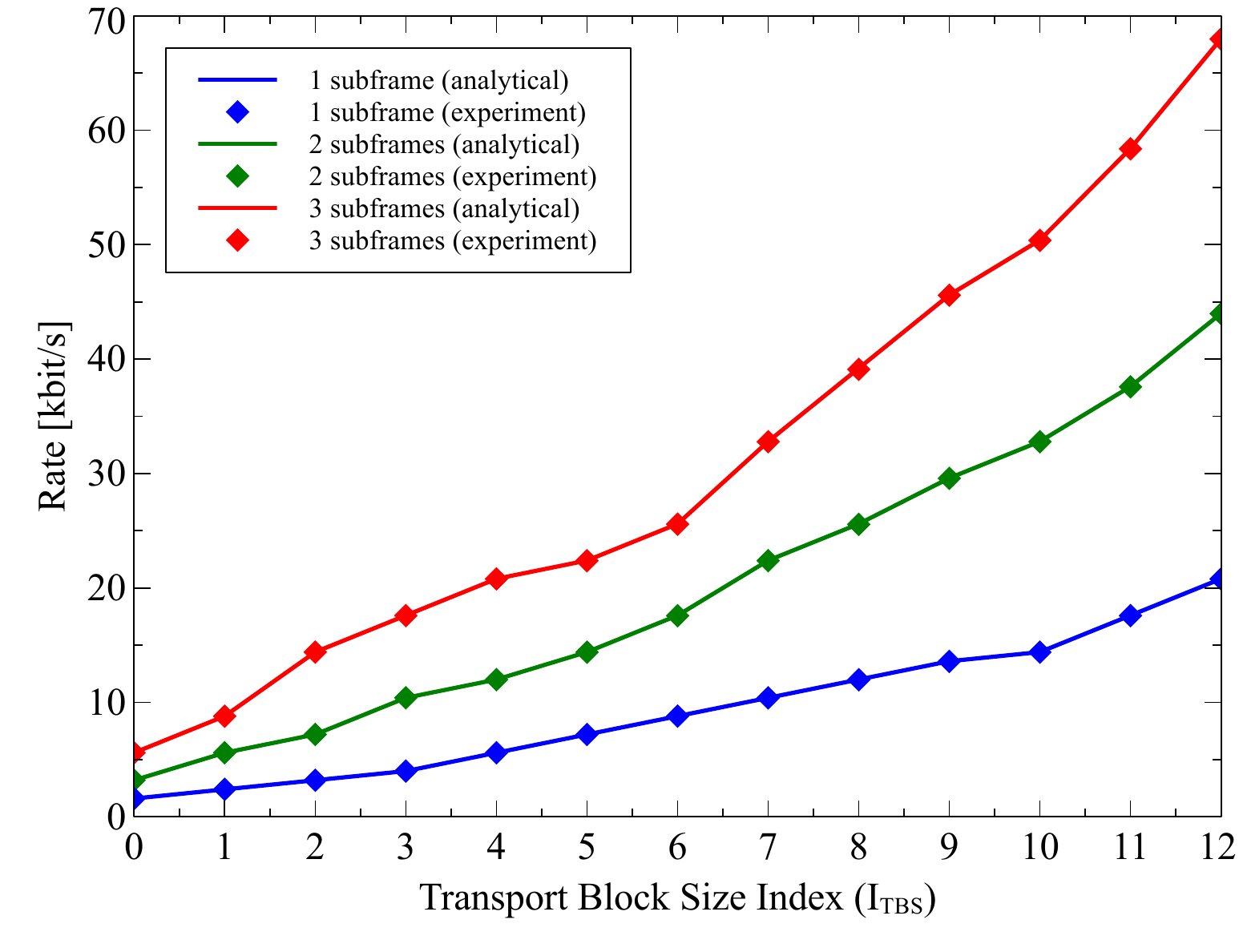}
		\caption{Comparison of theoretical and experimental DL rate (20~dB SNR).}
		\label{fig:tp_exp}
	\end{center}
\end{figure}

After the theoretical analysis, this section will now assess the performance of our prototype implementation
and compare it against them. In practice, the above stated results (e.g., 85 kbit/s DL throughput) can only
be achieved if the channel conditions between eNB and UE are excellent and the eNB is actually able to schedule
transmissions for a specific UE using the minimum number of subframes to, e.g., transmit 680~bit of data in
8~subframes only. Using a single NB-IoT downlink carrier, however, this is difficult to achieve because the
so-called anchor carrier requires specific subframes to be filled with periodic information, such as
synchronization signals (i.e. NPSS and NSSS) as well as system information messages (i.e., MIB and SIBs).
Those signals will always have higher priority as scheduling grants and user data carried over NPDCCH or
NPDSCH, respectively.
Our basic NB-IoT eNB example therefore uses a simple static resource allocation scheme in which a NPDCCH
containing a DL grant is transmitted in subframe 1 of each frame and the corresponding NPDSCH from
subframe 6 onwards in the same frame. Using this configuration allows us to transmit 680 Bits of data in
subframe 6, 7 and 8 of each frame, i.e., every 10ms. Hence, using this resource allocation scheme, the maximum
achievable DL rate is 68 kbit/s.
Figure \ref{fig:tp_exp} shows the results of an experiment to assess the maximum achievable DL rate
of our prototype implementation. The experiment was carried out under favourable radio conditions with an
SNR of approximately 20~dB. We ran the experiment with three different NPDSCH configurations, i.e., with one,
two, and three subframes. For each configuration, we also varied the TBS size from minimal to maximal
setting ($i_{tbs}$) value. In total, we ran 36 experiments. Each run lasted about one minute, until the
measured rate stabilized. From the plot it can be seen that in a high SNR regime with a NPDSCH bit error
rate of 0\%, the experimental results exactly match the theoretical calculation for each TBS. 
We are currently executing a set of simulations and experiments to evaluate the DL rate under lower SNR regimes.

\section{Exploring Commercial Deployments}
\label{sec:mwc}

Several network operators around the world have chosen to adopt NB-IoT for
MTC communication.
Vodafone, for example, began to roll out their network in Spain in late January 2017,
starting in Madrid and Valencia, but already announcing the expansion to other Spanish
cities, such as Barcelona, Bilbao, Málaga and Seville \cite{vodafone-commercial-launch}.
Only days later, on February 1st, we discovered a live NB-IoT carrier in downtown
Barcelona in Vodafone's 800 MHz frequency band, which has been used from
then on to test and validate standard conformance of our implementation.

\begin{figure}
	\begin{center}
		\includegraphics[width=1.0\columnwidth]{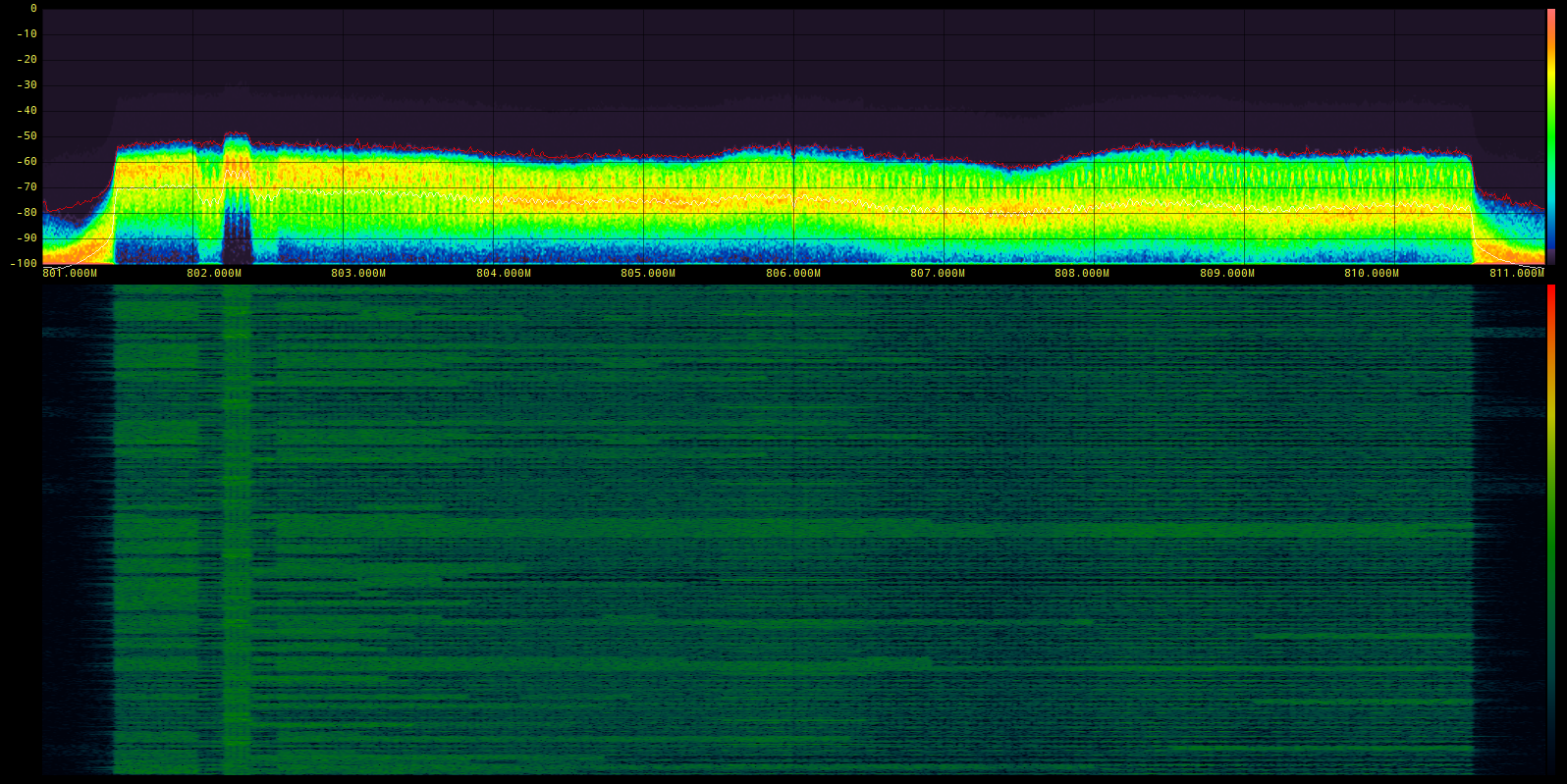}
		\caption{Waterfall plot of the 10 MHz Vodafone cell at 806 MHz in Barcelona, Spain with the NB-IoT carrier
			clearly visible in the left part of the signal.}
		\label{nbiot-full-bw}
	\end{center}	
\end{figure}

Figure \ref{nbiot-full-bw} shows a power spectral density and waterfall plot of the
aforementioned Vodafone LTE band centered at 806~MHz over the full cell bandwidth of 10~MHz.
The NB-IoT carrier is clearly visible on the left-hand side of the spectrum. It is
transmitted with slightly higher power than the rest of the LTE signal.
Because the NB-IoT carrier is embedded in and uses resources of an existing LTE cell,
it is said to operate in \textit{in-band} mode.

During Mobile World Congress~(MWC)~2017, SRS showcased a fully functioning DL receiver
and decoder for NB-IoT. This demonstration not only decoded the live transmission of the Vodafone
cell but also displayed various information about the signal, such as the Carrier Frequency Offset~(CFO), 
Sample Frequency Offset~(SFO) and Block Error Rate~(BER) of the NPDSCH.

\begin{figure}
	\begin{center}
		\includegraphics[width=\columnwidth]{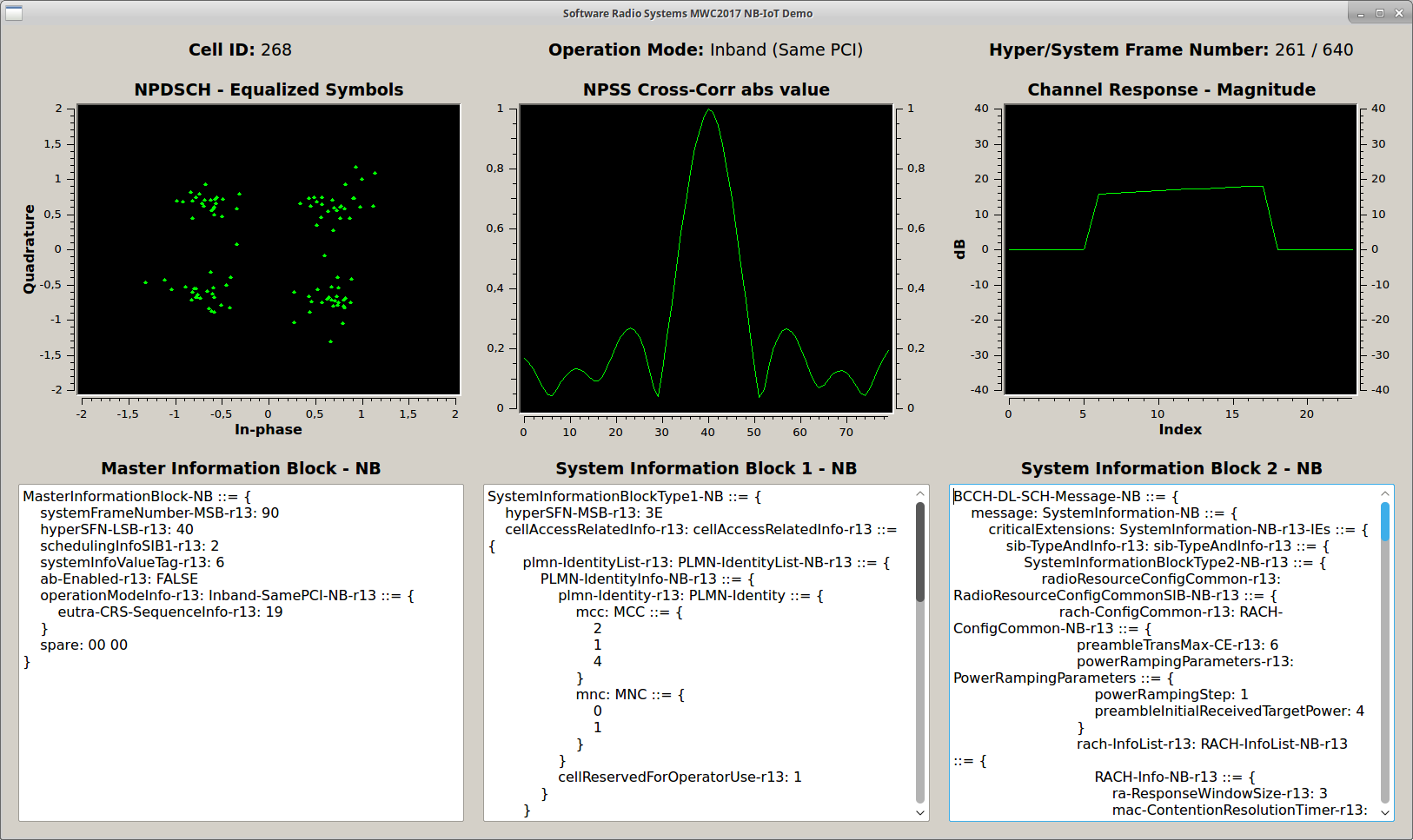}
	\end{center}
	\caption{Screenshot of the NB-IoT UE shown at MWC2017.}
	\label{nbiot-mwc-demo}
\end{figure}

Figure \ref{nbiot-mwc-demo} shows the graphical user interface of the UE displaying the constellation
diagram of NPDSCH~(upper left image), the correlation peak of the NPSS~(upper middle image) and the channel response~(upper right image). The UE also shows the content of MIB, SIB1 and SIB2~(bottom three text fields in Figure \ref{nbiot-mwc-demo}).

\section{Conclusion}
\label{sec:conclusion}

With the finalization of Release 13 of the LTE Advanced Pro specification
in June 2016, 3GPP has provided a new radio access technology based on LTE
to address the requirements and challenges of the Internet of Things.

In this paper, we have summarized our efforts in developing a software
implementation containing all relevant DL and UL channels and signals
defined in the standard as well as example applications that facilitate
testing and experimenting with this new technology.

Furthermore, the paper also described experiments carried out to assess the
performance of the implementation and compares the obtained results with our analytical model.
Our analysis showed that the theoretical peak data rate of 226~kbit/s in
the DL can’t be achieved in practice due to the control overhead
associated with any transmission. In fact, the maximum achievable DL
data rate for a single UE is 85~kbit/s in unacknowledged mode and
32.38~kbit/s in acknowledged mode, which can only be achieved with the
highest modulation and coding scheme, using the lowest number of subframes
and no repetitions.
The maximum UL rate of 68~kbit/s achieved in our experiments can be viewed as an
realistic upper limit in single carrier deployments.
The 20\% reduction compared to the theoretical maximum
can be well explained by the overhead caused by the periodic DL transmissions,
i.e., synchronization signals and system information messages.

\section*{Acknowledgments}
The research and development leading to these results has received
funding from the European Research Council under the European
Community Seventh Framework Program (FP7/2014-2017) grant
agreement 612050 (FLEX project).

\bibliographystyle{IEEEtran}
\bibliography{src/citations}

\begin{thebibliography}{1}
\providecommand{\url}[1]{#1}
\csname url@samestyle\endcsname
\providecommand{\newblock}{\relax}
\providecommand{\bibinfo}[2]{#2}
\providecommand{\BIBentrySTDinterwordspacing}{\spaceskip=0pt\relax}
\providecommand{\BIBentryALTinterwordstretchfactor}{4}
\providecommand{\BIBentryALTinterwordspacing}{\spaceskip=\fontdimen2\font plus
\BIBentryALTinterwordstretchfactor\fontdimen3\font minus
  \fontdimen4\font\relax}
\providecommand{\BIBforeignlanguage}[2]{{%
\expandafter\ifx\csname l@#1\endcsname\relax
\typeout{** WARNING: IEEEtran.bst: No hyphenation pattern has been}%
\typeout{** loaded for the language `#1'. Using the pattern for}%
\typeout{** the default language instead.}%
\else
\language=\csname l@#1\endcsname
\fi
#2}}
\providecommand{\BIBdecl}{\relax}
\BIBdecl

\bibitem{Wang2016}
Y.~P.~E. Wang, X.~Lin, A.~Adhikary, A.~Grovlen, Y.~Sui, Y.~Blankenship,
  J.~Bergman, and H.~S. Razaghi, ``{A Primer on 3GPP Narrowband Internet of
  Things},'' \emph{IEEE Communications Magazine}, vol.~55, no.~3, March 2017.

\bibitem{rohde-paper-nbiot}
{Rohde \& Schwarz}, ``{Narrowband Internet of Things - Whitepaper},'' 2016,
  available at \url{http://www.rohde-schwarz.com/appnote/1MA266}.

\bibitem{ts36211}
ETSI, ``{TS 136 211: LTE: Evolved Universal Terrestrial Radio Access (E-UTRA):
  Physical channels and modulation},'' 2016.

\bibitem{srslte}
{Software Radio Systems}, ``{srsLTE: Open Source LTE},'' available at
  \url{https://github.com/srsLTE}.

\bibitem{vodafone-commercial-launch}
{Vodafone}, ``{Commercial launch of NB-IoT in Vodafone Spain},'' 2017,
  available at
  \url{http://www.vodafone.com/content/index/what/technology-blog/nbiot-commercial-launch-spain.html}.

\end{thebibliography}

\end{document}